# Study of the effect of electromagnetic damping force on a magnet oscillating near a non-ferromagnetic conducting plate

Sanjoy Kumar Pal[1], Soumen Sarkar[2], Pradipta Panchadhyayee[3,4*], and Debapriyo Syam[5]

[1]Anandapur H.S. School, Anandapur, PaschimMedinipur, West Bengal, India
[2]Karui P.C. High School, Hooghly, West Bengal, India
[3]Department of Physics (UG & PG), Prabhat Kumar College, Contai, PurbaMedinipur, India
[4]Institute of Astronomy, Space and Earth Science, Kolkata -700054, W. B., India
[5]Kalpana Chawla Centre for Space and Nanosciences, Kolkata -. 700010, India
[*]E-mail: ppcontai@gmail.com

**Abstract**

We have designed an experiment that involves studying the effects of a conducting plate on the motion of an oscillating disc magnet. We have employed the video analysis method by ‘Tracker’ software to investigate the variation of electromagnetic damping coefficient with distance between the plate and the magnet. This experiment can indeed serve as a valuable educational tool for undergraduate students, covering topics such as damped oscillation, electromagnetic damping, Lenz's law, and eddy currents.

## Introduction

In practice, damping is a crucial aspect of controlling the motion of objects, and it involves the dissipation of energy to reduce or control oscillations, vibrations, or other undesired motions. Damping is commonly used in various systems to enhance stability, reduce noise, and prevent damage. The damping techniques include (i) Conventional frictional damping, (ii) Air friction (i.e. viscous) damping, (iii) Fluid friction (viscous) damping, and (iv) Electromagnetic (EM) damping.

In EM damping, the basic operating principle is as follows: When a moving metallic object passes through a magnetic field, it induces eddy currents (loops of electric current) in the object. These currents, in turn, create the corresponding magnetic field. Interaction between the magnetic field due to the induced current and the disc magnet results in slowing down (damping) of the object's motion. This method is often used in precision instruments and systems where physical contact with damping materials might be undesirable. The advantage of EM damping is that it can provide effective damping without introducing wear and tear associated with physical contact damping methods. However, designing and implementing electromagnetic damping systems can be complex and requires careful consideration of the specific application requirements.

Addressing conceptual challenges related to electromagnetic damping (also called braking) several researchers reported experiments [1-22] on Lenz's law and eddy currents, which are the building blocks of the related theory. Some of them have presented the experiments with a view to demonstrating Lenz's law [1-9]. The role of eddy currents in the damped falling motion of magnets inside conducting tubes [10-16] and in magnetic levitation [17] has also been investigated. Some other researchers have conducted detailed analyses of EM damping [18-21]. Different physical quantities like displacement, velocity, and acceleration [22] and current waveforms [23-28] were studied for magnetically damped oscillations. Inspired by these works, we have designed a demonstration experiment on the dependence of the EM damping coefficient on distance between magnet and a metal plate. It can also be considered as a practical application of Newton's third law. Our experimental arrangement consisted of a disc magnet attached to the lower end of a wooden strip hung from a stout pin and a thin aluminium plate. A pair of slide-callipers, a smartphone and a computer served as data acquisition and analysis devices. We believe that this experimental arrangement would be found interesting by undergraduate students and their teachers.

## Theoretical Framework

When a magnet oscillates near a conductor, it induces ring currents in the latter. This is, of course, the phenomenon of electromagnetic induction. The induced current gives rise to a magnetic field that opposes the motion of the magnet. If the magnet is suspended as a pendulum in front of the conductor, the opposing force produces the damping of the oscillatory motion (over and above the damping due to air resistance). Fig.1a shows a schematic drawing of the experimental arrangement and Fig. 1b is drawn to facilitate the understanding of the calculations.

The magnetic field produced by the disc magnet at the centre *P* of the shadow is given by

$$\vec{B} = -\frac{\mu_0}{4\pi} \cdot \frac{2M}{r_P^3} \hat{k} = -B\hat{k} \qquad (1)$$

where $B = \frac{\mu_0}{4\pi} \cdot \frac{2M}{r_P^3}$ ($\hat{i}$, $\hat{j}$, $\hat{k}$ represent the unit vectors in the x, y and z directions, respectively). $M$ is the magnetic dipole moment of the disc magnet, and $r_P = \sqrt{d^2 + a^2}$. We assume that $d \le a$. Here, $d$ is the distance between the centre of gravity of the magnet and the aluminium plate and the radius of the disc magnet is $a$.

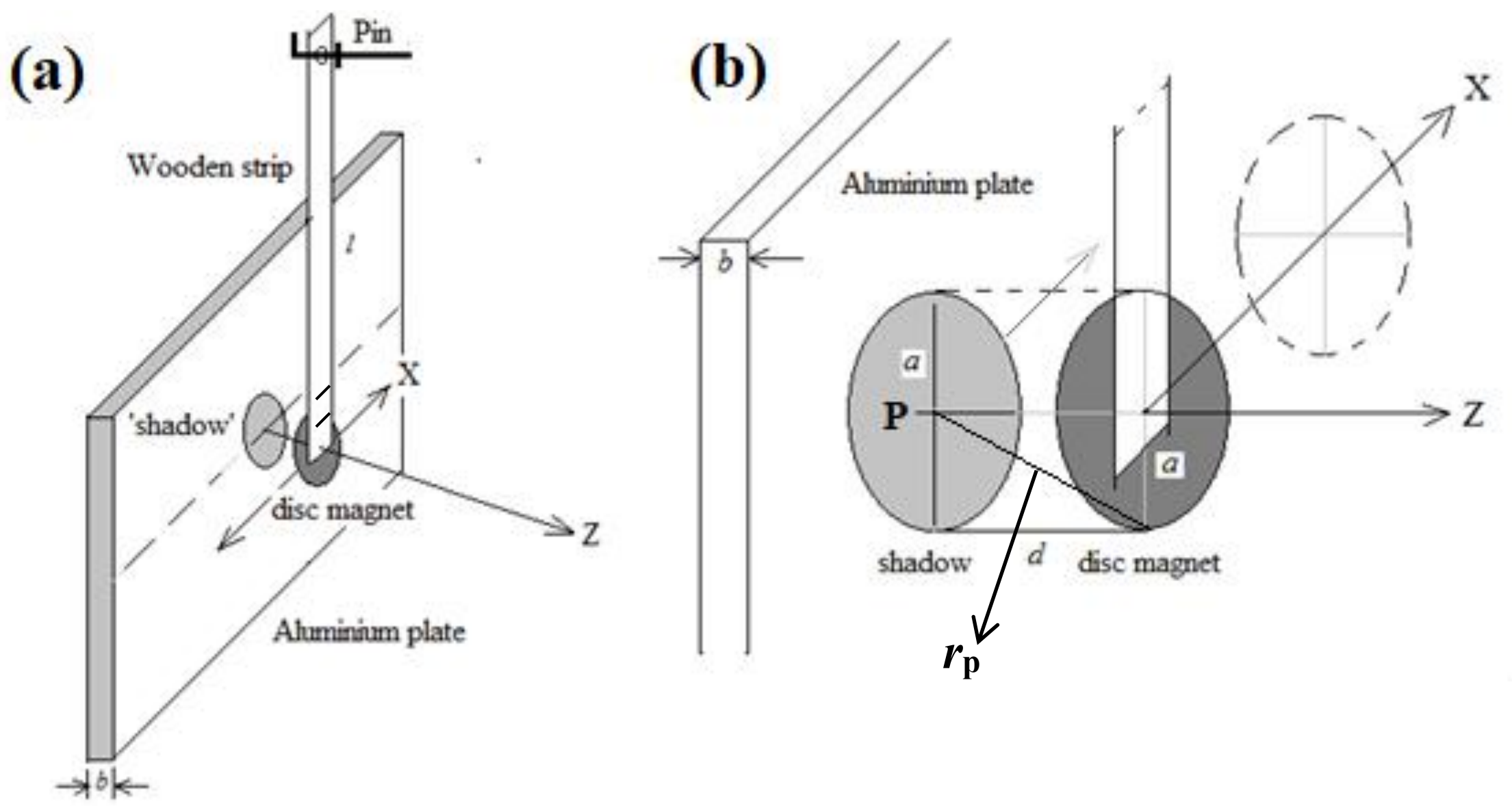

Fig. 1: (a) – Schematic diagram of the experimental setup. (b) – This figure is drawn to facilitate the understanding of the calculations.

*We assume that the magnet oscillates in the magnetic east-west direction and that it never crosses the edges of the aluminium plate.*

If $\vec{v}$ stands for the relative velocity of the aluminium plate with respect to the magnet, then the effective value of the electric field at $P$ is

$$\vec{E}_{eff} = \vec{v} \times \vec{B} = -\left(\frac{dx}{dt}\right)\hat{i} \times \left(-B\hat{k}\right) = -\left(\frac{dx}{dt}\right)B\hat{j}. \tag{2}$$

where $x$ is the displacement of the centre of the bob (disc magnet).

The induced current density is $\vec{j} = \sigma \vec{E}_{eff}$, where $\sigma$ is the electrical conductivity of aluminium.

We note that a cylindrical region appears under the shadow with the volume which is a fraction ($f$) of the volume element, $\pi a^2 b$ ($b$ = thickness of aluminium plate). More generally, one may assume, $f \approx 1$. The force on the volume element is

$$\vec{F} = \left(\pi a^2 b\vec{j}\right) \times \vec{B} = -\pi a^2 b\sigma\left(\frac{dx}{dt}\right)B\,\hat{j} \times (-B\hat{k}) = \pi a^2 b\sigma\left(\frac{dx}{dt}\right)B^2\hat{i} \tag{3}$$

The force of reaction on the magnet, which produces the braking effect, is, therefore,

$$\vec{F}_{mag} = -\pi a^2 b\sigma\left(\frac{dx}{dt}\right)B^2\hat{i} \equiv -\beta\left(\frac{dx}{dt}\right)\hat{i}. \tag{4}$$

Here, the expression $\pi a^2 b\sigma B^2$ is written as $\beta$.

Ignoring air resistance, the equation of motion of the magnet when suspended by a weightless strip, is:

$$m\frac{d^2x}{dt^2}+\beta\cdot\frac{dx}{dt}+\frac{mgx}{l}=0, \tag{5}$$

where $m$ is the mass of the magnet, $g$ is strength of the Earth's gravitational field and $l$ is the length of the suspension strip.

If the mass ($m_s$) of the suspension strip is not negligible, then it is better to first write the equation of motion in terms of the angular displacement $\theta$ of the suspended system from its equilibrium position. This equation is

$$\left(ml^2+\frac{m_s l^2}{3}\right)\frac{d^2\theta}{dt^2}+\beta l^2\frac{d\theta}{dt}+\left(ml+m_s\frac{l}{2}\right)g\sin\theta=0$$

Or

$$\left(m+\frac{m_s}{3}\right)l\frac{d^2\theta}{dt^2}+\beta l\frac{d\theta}{dt}+\left(m+\frac{m_s}{2}\right)g\sin\theta=0, \tag{6}$$

where we have assumed that the strip has a uniform value of mass per unit length.

When $\theta$ is small, we may write, $sin\theta=\frac{x}{l}\approx\theta$. Consequently, the equation of motion becomes

$$(m+\frac{m_s}{3})\cdot\frac{d^2x}{dt^2}+\beta\cdot\frac{dx}{dt}+\frac{(m+\frac{m_s}{2})gx}{l}=0$$

We may rewrite this equation as $\frac{d^2x}{dt^2}+2\alpha\cdot\frac{dx}{dt}+\omega_0^2x=0$ (7)

with $2\alpha=\frac{\beta}{m+\frac{m_s}{3}}$; $\omega_0^2=\frac{m+\frac{m_s}{2}}{m+\frac{m_s}{3}}\cdot\frac{g}{l}$.

The solution of Eq. (7) can be written as $x=(C\cos\omega t+D\sin\omega t)e^{-\alpha t}$,
where $\omega=\sqrt{\omega_0^2-\alpha^2}$.

Recalling the value of $\beta$, we note that $\alpha=\frac{3}{2}\cdot\frac{\pi a^2 b\sigma B^2}{3m+m_s}$ (with $B=\frac{\mu_0}{4\pi}\cdot\frac{2M}{r_P^3}$). (8)

When air resistance is also taken into account, the total damping coefficient is

$$\alpha=\alpha_{em}+\alpha_{air}\ , \tag{9}$$

where we have assumed that the air resistance is proportional to the velocity of the disc.
Here,

$$\alpha_{em}=\frac{3}{2}\cdot\frac{\beta}{3m+m_s} \tag{10}$$

As $\alpha_{em}$ falls off very quickly with distance ($d$) of the magnet from the aluminium sheet, so, for large $d$, $\alpha\approx\alpha_{air}$.
So, the amplitude ($R$) of oscillation falls off with time ($t$) as $\sqrt{C^2+D^2}e^{-\alpha_{air}t}$.
Thus, $\ln R\approx-\alpha_{air}t+\text{constant}$. (11)
From a graph of $\ln(R)$ against $t$, the value of $\alpha_{air}$ can be determined.

On the other hand, for small $d$, the graph of $\ln R$ versus $t$ has the equation

$$\ln R \cong -(\alpha_{em} + \alpha_{air})t + \text{constant.} \qquad (12)$$

Knowing the value of $\alpha_{air}$, the value $\alpha_{em}$ can, therefore, be readily ascertained following Eq. (9).

From the Eq. (10) for the EM damping coefficient we have

$$\alpha_{em}{}^{1/3} \propto \frac{1}{r_P{}^2}$$

or $$\alpha_{em}{}^{-1/3} \propto r_P{}^2 \ (= \ d^2 + a^2) \qquad (13)$$

So, we expect that the experimental data will yield:

$$\alpha_{em}{}^{-1/3} \propto \ d^2 + \text{ constant.} \qquad (14)$$

## Experiment and Results

Fig. 1a shows a wooden strip (length 94.0 cm, breadth 2.20 cm, depth 0.80 cm) suspended from a stout iron pin. A small ferrite disc magnet (diameter and width being 1.80 cm and 0.48 cm, respectively) is attached at the lower end of the strip. The magnet is oriented in the north-south direction of the earth's magnetic field and is free to oscillate in the east-west direction. In this case, the axis of the magnetic dipole (associated with the disc magnet) remains parallel to the horizontal component of the geomagnetic field. Thus, no torque due to the geomagnetic field acts on the disc. So, the disc moves in the vertical plane defined by the east-west direction. In front of the magnet, an aluminium sheet is placed vertically so that the distance ($d$) between the magnet and the aluminium plate always remains constant during the oscillations (amplitude nearly 1 cm) of the magnet. In this connection, we note that the assumption of uniform magnetic field over the shadow region is certainly an approximation. We have measured the distance between the magnet and aluminium plate by a brass made slide-callipers of 0.01 cm accuracy.

A smartphone in video mode is placed on a small stand in front of the lower end of the oscillator. Two small dashes are marked on the strip for calibration. Now, the oscillator is allowed to oscillate, and the smartphone records the oscillations with a frame rate of 60 frames per second. The recorded video files are transferred to a laptop and analyzed using the 'Tracker' software [29].

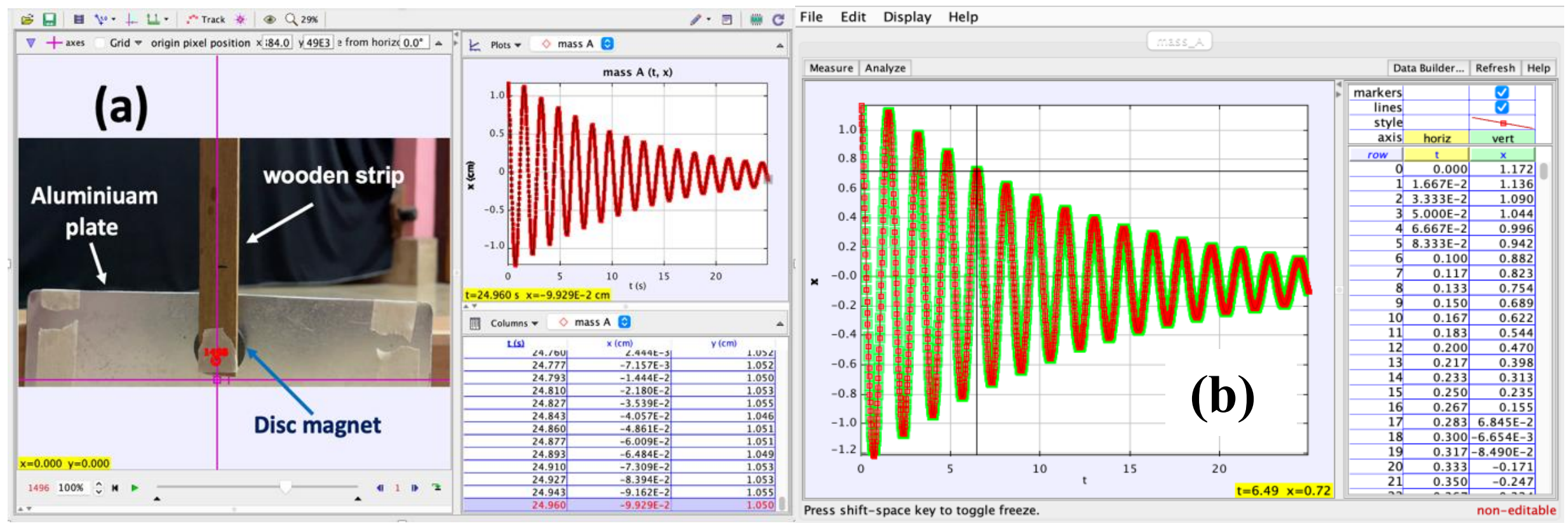

Fig. 2: (a) - Study of the recorded video using 'Tracker' software and plot of damping profile, (b) – Zoomed-in version of the damping profile.

By opening the video in the 'Tracker' application, we have calibrated the coordinates by placing the value of the distance between the two dots drawn on the wooden strip in the desired locations of the software. Next, we have selected the appropriate coordinates by setting the origin at the equilibrium position of the pendulum. By locating the lower black dot and using autotrack to follow the dot, the time vs displacement ($x$-direction) graph is obtained. Manually identifying the coordinates of the maximum displacement (amplitude), we have recorded the time versus amplitude data. Subsequently, the graph of the natural logarithm of amplitude ($\ln R$) with time ($t$) is plotted to determine the slope of the graph using linear fitting. We have computed the total damping coefficient ($\alpha$) from the slope. In order to determine the damping coefficient in air we have removed the aluminium plate from behind the magnet. The same procedure has been repeated using a wooden board as a substitute for the aluminium plate and the value of the damping coefficient in air ($\alpha_{air}$) is obtained as 0.0029 $s^{-1}$. As we have mentioned above, $\alpha$ is the sum of the damping coefficient in air ($\alpha_{air}$) and the EM damping coefficient ($\alpha_{em}$) under open-air conditions. We determine the value of $\alpha_{em}$ using the relation (Equation (9)) between $\alpha$, $\alpha_{air}$, and $\alpha_{em}$. We have tabulated readings (see Table I) for six different distances ($d$) measured from the aluminium plate to the centre of gravity of the magnet. For a particular value of $d$, we have repeated the experiment five times. The $\ln R$ vs $t$ graph for determining $\alpha_{air}$ and one representative $\ln R$ vs $t$ graph (when $d$ = 0.425 cm) for the computation of $\alpha_{em}$ are shown in Fig. 3a and Fig. 3b.

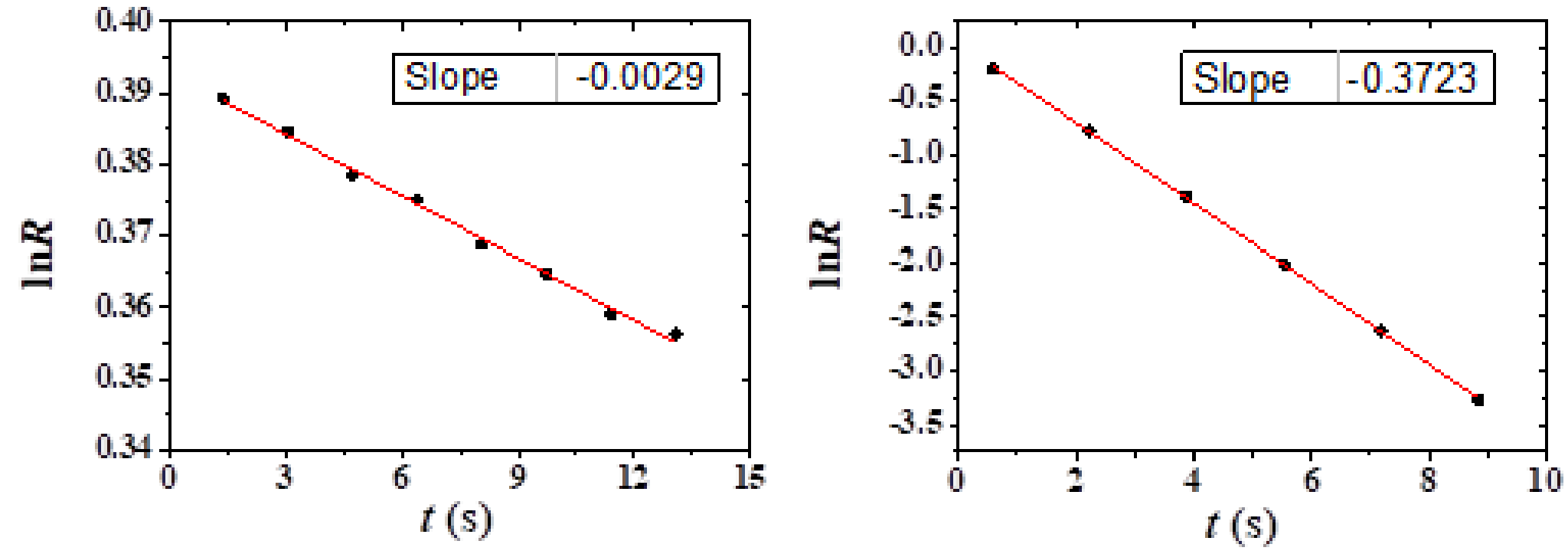

Fig. 3: (a) - $\ln R$ vs $t$ plot to determine $\alpha_{air}$, (b) $\alpha_{em}$ when $d$ = 0.425(0.43?) cm.

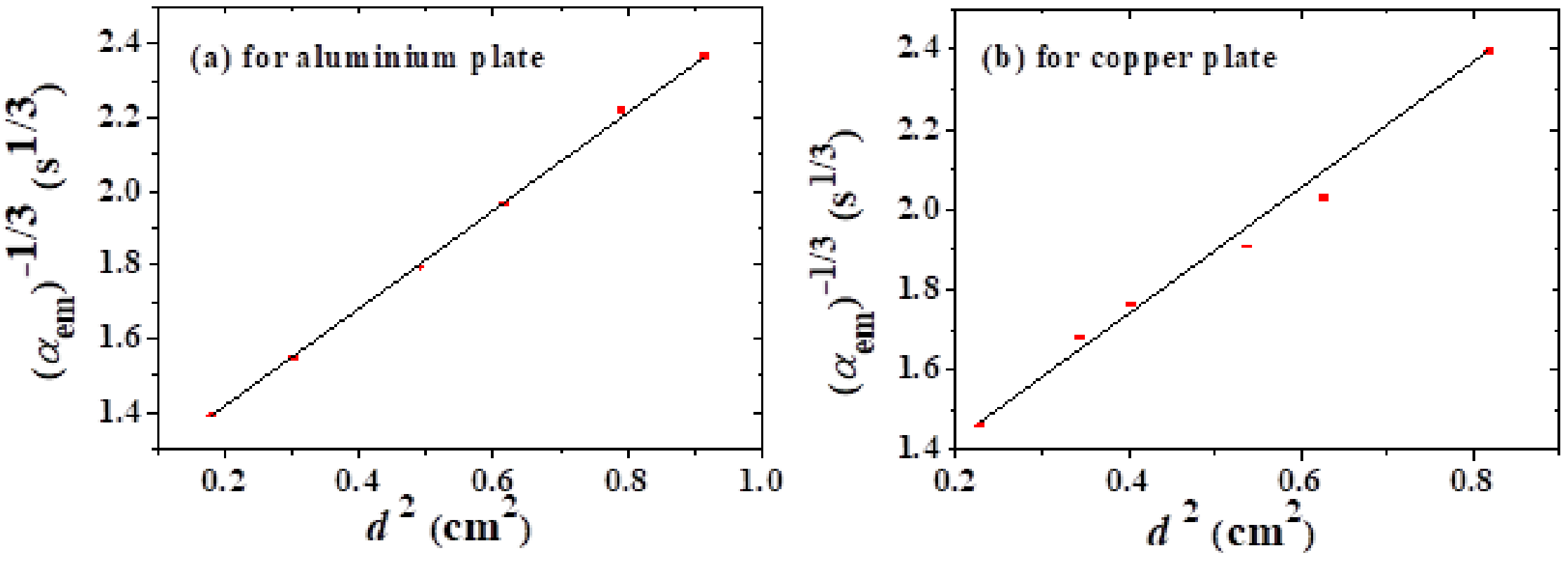

Fig. 4: $(\alpha_{em})^{-\frac{1}{3}}$ vs $d^2$ plots with error bars: (a)- for aluminum plate; (b)- for copper plate.

Table I
Table for distance ($d$) vs total damping coefficient ($\alpha$) and EM damping coefficient ($\alpha_{em}$)
(Note that the damping coefficient in air ($\alpha_{air}$) = 0.0029 $s^{-1}$)

| $d$ (cm) | $d^2$($cm^2$) | $\alpha$($s^{-1}$) | $\alpha_{em}$ ($s^{-1}$) | $(\alpha_{em})^{-\frac{1}{3}}(s^{\frac{1}{3}})$ | Average $\alpha_{em}$ ($s^{-1}$) |
|---|---|---|---|---|---|
| 0.42 | 0.1764 | 0.3722 | 0.3693 | 1.3938 | 0.3690 ±0.0074 ±0.0004 |
| | | 0.3713 | 0.3684 | 1.3950 | |
| | | 0.3723 | 0.3694 | 1.3937 | |
| | | 0.3708 | 0.3679 | 1.3956 | |
| | | 0.3731 | 0.3702 | 1.3927 | |
| 0.55 | 0.3025 | 0.2733 | 0.2704 | 1.5464 | 0.2700 ±0.0053 ±0.0004 |
| | | 0.2720 | 0.2691 | 1.5489 | |
| | | 0.2729 | 0.2700 | 1.5472 | |
| | | 0.2724 | 0.2695 | 1.5482 | |
| | | 0.2741 | 0.2712 | 1.5449 | |
| 0.70 | 0.4900 | 0.1758 | 0.1729 | 1.7950 | 0.1731 ±0.0032 ±0.0002 |
| | | 0.1755 | 0.1726 | 1.7960 | |
| | | 0.1766 | 0.1737 | 1.7922 | |
| | | 0.1763 | 0.1734 | 1.7933 | |
| | | 0.1756 | 0.1727 | 1.7957 | |
| 0.79 | 0.6141 | 0.1341 | 0.1312 | 1.9680 | 0.1315 ±0.0023 ±0.0001 |
| | | 0.1345 | 0.1316 | 1.9660 | |
| | | 0.1346 | 0.1317 | 1.9655 | |
| | | 0.1345 | 0.1316 | 1.9660 | |
| | | 0.1341 | 0.1312 | 1.9680 | |
| 0.89 | 0.7921 | 0.0939 | 0.0910 | 2.2232 | 0.0915 ±0.0020 ±0.0002 |
| | | 0.0952 | 0.0923 | 2.2128 | |
| | | 0.0943 | 0.0914 | 2.2200 | |
| | | 0.0939 | 0.0910 | 2.2232 | |
| | | 0.0946 | 0.0917 | 2.2176 | |
| 0.96 | 0.9216 | 0.0784 | 0.0755 | 2.3660 | 0.0755 ±0.0020 ±0.0001 |
| | | 0.0783 | 0.0754 | 2.3671 | |
| | | 0.0786 | 0.0757 | 2.3639 | |
| | | 0.0780 | 0.0751 | 2.3702 | |
| | | 0.0785 | 0.0756 | 2.3650 | |

Based on the data and results in Table I we have plotted Fig. 4(a) which clearly displays the linear variation of $(\alpha_{\text{em}})^{-\frac{1}{3}}$ with $d^2$, as expected from Eq. (13). To test the validity of the method, we have repeated the experiment using a copper plate instead of the aluminium plate, and the variation of $(\alpha_{\text{em}})^{-\frac{1}{3}}$ with $d^2$ is shown in the Fig. 4(b). A similar variation is observed for the copper plate, like that shown in Fig. 4(a) for the aluminium plate. The value of the constant term (0.87 $\text{cm}^2$ for aluminium and 0.73 $\text{cm}^2$ for copper) in Eq. (14) can be estimated from Fig. 4(a), and it nearly matches the actual value of $a^2$ (0.81 $\text{cm}^2$). In Table 1, for the aluminium plate, the last column shows the average values of $\alpha_{\text{em}}$ for the corresponding values of $d$ along with the instrumental uncertainty and statistical error. It is clearly noticeable that the contribution of instrumental uncertainty in computing $\alpha_{\text{em}}$ becomes less and gets a stable value with the increase in $d$.

**Conclusion:**

In this paper, we describe an experiment to explore how the presence of a conducting plate like an aluminium sheet affects the motion of an oscillating disc magnet, explicitly focusing on electromagnetic interactions. Using video analysis with 'Tracker' software, we have added a modern and quantitative aspect to our investigation. We have noted the decrease in the oscillation amplitude of the magnet-bearing pendulum due to air damping and EM damping using Tracker software. First, we have determined the value of the damping coefficient in air ($\alpha_{\text{air}}$) in the presence of the wooden board. Finally, we have verified the theoretically anticipated variation of the EM damping coefficient for different distances between the aluminium plate and the oscillating magnet. To establish the robustness of the method, we have repeated the experiment using a copper plate and obtained a variation of the EM damping coefficient with the distance between the metallic plate and the magnet that is similar to the one shown for the aluminium plate. This experiment not only offers a hands-on approach but also covers important concepts such as damped oscillation, electromagnetic damping, Lenz's law, and eddy currents.

**Acknowledgement:** We thankfully acknowledge Dr. Subhash Chandra Samanta and Dr. Syed Minhaz Hossain for their valuable suggestions during the experiment.